\documentclass[10pt, aps,pra,twocolumn,superscriptaddress,floatfix,nofootinbib,amsmath,amssymb]{revtex4-2}

\usepackage{graphicx}
\usepackage[normalem]{ulem}
\usepackage{bm}
\usepackage{braket}

\usepackage{xcolor}
\usepackage[
  colorlinks=true,
  linkcolor=blue,
  citecolor=blue,
  urlcolor=blue,
  breaklinks=true
]{hyperref}
\usepackage[capitalise]{cleveref}

\begin{document}

\author{Johannes Christmann}
\thanks{equal contribution}
\affiliation{Departement of Physics, ETH Zürich, Zürich, Switzerland}

\author{Petr Ivashkov }
\thanks{equal contribution}
\affiliation{Department of Information Technology and Electrical Engineering, ETH Zürich, Zürich, Switzerland}

\author{Mattia Chiurco }
\affiliation{Departement of Physics, ETH Zürich, Zürich, Switzerland}

\author{Guglielmo Mazzola}
\email[]{guglielmo.mazzola@uzh.ch}
\affiliation{Department of Astrophysics, University of Zürich, Winterthurerstrasse 190, 8057 Zürich, Switzerland}

\date{today}

\title{From quantum-enhanced to quantum-inspired Monte Carlo}
\date{\today}

\begin{abstract}
We perform a comprehensive analysis of the quantum-enhanced Monte Carlo method [\textit{Nature}, 619, 282 (2023)], aimed at identifying the optimal working point of the algorithm.
We observe an optimal mixing Hamiltonian strength and analyze the scaling of the total evolution time with the size of the system. We also explore extensions of the circuit, including the use of time-dependent Hamiltonians and reverse digitized annealing.
Additionally, we propose that classical, approximate quantum simulators can be used for the proposal step instead of the original real-hardware implementation.
We observe that tensor-network simulators, even with unconverged settings, can maintain a scaling advantage over standard classical samplers. This may extend the utility of quantum enhanced Monte Carlo as a quantum-inspired algorithm, even before the deployment of large-scale quantum hardware.
\end{abstract}

\maketitle

\section{Introduction} 

Monte Carlo methods are widespread across multiple disciplines, extending beyond the traditional boundaries of natural science. Besides predicting phase diagrams of materials, lattice models, and chemical reactions, they are widely used in engineering, machine learning, and finance, to name just a few. The development of better sampling schemes is therefore a central technological challenge~\cite{binder2012monte,bishop2023deep,kirkpatrick1983optimization,allen2017computer}.

Very recently, with the advent of quantum computation~\cite{nielsen_chuang_2010}, efforts have been made to devise quantum algorithms that could speed up the sampling tasks of classical lattice models~\cite{mazzola2024quantum}.
Quantum walks were proposed decades ago~\cite{venegas2012quantum}, but they require a long coherent evolution, and it is not clear whether the quadratic speed-up they offer is enough to provide a real practical advantage~\cite{lemieux2020efficient,babbush:2021}.
More recently, it has been proposed that wavefunction collapses could be a powerful computational resource for uncorrelated configurations sampling in physical models~\cite{mazzola2021sampling}.
This general idea was later formalized in the quantum-enhanced Markov chain Monte Carlo (QEMC) algorithm, which features a hybrid strategy~\cite{layden2023quantum}: The proposal step of the Markov chain is performed using quantum hardware, while the acceptance step, which requires evaluating the cost-function difference between the current and proposed configurations, is done classically.
The quantum update is realized using a Hamiltonian evolution circuit, implemented through the Trotter algorithm.
The main advantage of this method is that coherent evolution is not required throughout the entire Markov chain, making it a plausible candidate for near-term quantum advantage. Moreover, the method displays an empirical superquadratic advantage over the best local update Metropolis scheme for average-case disordered instances, although this evidence is currently limited to lattice sizes of up to about 20.
The algorithm has been demonstrated on real quantum hardware, displaying an interesting resilience to hardware noise. However, the evidence for quantum advantage is empirical, and a clear understanding of why the method works is still lacking. This is, however, common in the field of Monte Carlo, where the efficiency of sampling schemes is ultimately proven numerically.

Multiple recent studies have been dedicated to better understanding the algorithm in exactly solvable, specific models~\cite{orfi2024bounding} or the parameter's limits~\cite{orfi2024quantum}, proposing alternative circuit implementations~\cite{nakano2024markov}, extensions to continuous models~\cite{lockwood2024quantum}, or further exploring parallelization possibilities~\cite{ferguson2024quantum}.

In this paper we focus on two major points. The first concerns the sensitivity of the quantum speed-up with respect to the QEMC algorithm's parameters, namely the strength of the mixing Hamiltonian and the total evolution time (see Sec.~\ref{s:QEMC}). This numerical analysis is important to determine whether the algorithm requires excessive instance-dependent fine-tuning, which would make it impractical, and whether there is a hidden relationship between the simulation time and the system size. Clearly, if the depth of the circuit needs to grow exponentially with the system size, this would diminish the quantum advantage.

The second, conceptually novel, point is to explore whether a quantum-inspired version of the algorithm, running on classical machines, can also be effective. While classical methods cannot exactly simulate quantum dynamics, QEMC uses the Hamiltonian simulation circuit only as an update proposal generator. It remains unclear whether algorithmic errors can significantly impact the overall quantum speedup.
For instance, it is known that running simulated quantum annealing~\cite{santoro2002theory,albash2018adiabatic} (via path-integral Monte Carlo simulations) in an unconverged, unphysical setting can actually be beneficial if this method is used for optimization. In such cases, large imaginary-time Trotter steps~\cite{heim2015quantum} or open imaginary-time boundary conditions~\cite{isakov2016understanding,mazzola2017quantum} improve the probability of escaping local minima. Therefore, it is not ruled out that using the Hamiltonian circuit with a larger Trotter step, or even more radically, employing a classical approximation of quantum dynamics within QEMC, could preserve, or possibly boost, its efficiency.

The paper is organized as follows. In Sec.~\ref{s:QEMC} we introduce the original quantum-enhanced Monte Carlo algorithm and state the requirements fo practical quantum advantage. In Sec.~\ref{s:optparam} we identify the optimal working point of the algorithm. In Sec.~\ref{s:optsched} we test different circuits for the proposal step. In Sec.~\ref{s:tns} we introduce the quantum-inspired version of the technique. In Sec.~\ref{s:conclu} we summarize the work, discuss our conclusions, and present an outlook for future work.

\section{Quantum enhanced Markov chain Monte Carlo}
\label{s:QEMC}

\subsection{Spin glasses}
Let $H_c(s)$ be a classical Ising model Hamiltonian and $s \in \{\pm1\}^n$ be a classical state in the configuration space of a system of $n$ spin variables. An Ising model is defined by coefficients $\{J_{ij}\}$ and $\{h_i\}$, called couplings and fields, respectively. A configuration $s$ has energy
\begin{equation}
    H_c(\textbf{$s$}) = - \frac{1}{2}\sum_{i,j=1}^{n} J_{ij} s_i s_j - \sum_{i=1}^{n} h_i s_i.
\end{equation}
In this paper, we consider a fully connected graph such that all $J_{ij}$ are nonzero. We focus on the fully connected model, as it is more challenging. The corresponding Boltzmann distribution $\pi(s)$ at temperature $T = 1/\beta$ is given by
\begin{equation}
    \pi(s) = \frac{e^{-\beta H_c(s)}}{\mathcal{Z}}
\end{equation}
where $\mathcal{Z} = \sum_{s}{e^{-\beta H_c(s)}}$ is the normalization constant. Usually, computing the Boltzmann distribution explicitly is computationally intractable due to an exponentially large summation in the normalization constant. 
Ising models with random ${J_{ij}}$ and ${h_i}$, also known as spin glasses, typically exhibit rugged energy landscapes with many local minima~\cite{mezard1987spin,lucas2014ising,Edwards_1975,barahona1982computational}.
While this type of Hamiltonian may seem artificial, the model is ubiquitous across many fields of science and engineering, from materials science and optimization to biological networks~\cite{Binder_glass,de2016simple}. Historically, disordered spin glasses have been considered an ideal testbed for developing and testing novel classical~\cite{kirkpatrick1983optimization,santoro2002theory, zhu2015efficient, barzegar:2018, houdayer2001cluster} and quantum algorithms and hardware~\cite{johnson2011quantum, boixo2014evidence, ronnow2014defining, baldassi2018efficiency, pagano2020quantum, layden2023quantum}.

\subsection{Markov chain Monte Carlo}
Markov chain Monte Carlo (MCMC) is the most widely used technique for sampling from intractable Boltzmann distributions~\cite{robert2011short,metropolis1953equation}. It avoids explicit computation of $\pi(s)$ through an efficient two-step process. Initially, starting from a spin configuration $s_0$, a new configuration $s$ is proposed with a probability $Q(s|s_0)$. Subsequently, the proposed configuration is accepted with probability $A(s|s_0)$. These two steps form a Markov chain, whereby the transition from $s_0$ to $s$ occurs with a probability
\begin{equation}
P(s_0 \rightarrow s) = Q(s|s_0) A(s|s_0),
\end{equation}
where $P$ represents a stochastic transition matrix. A Markov chain that meets the criteria of irreducibility, aperiodicity, and the detailed balance condition will converge to a stationary distribution whereby the specific choices of $Q(s|s_0)$ and $A(s|s_0)$ determine the desired stationary distribution~\cite{levin2017markov}. Among the most commonly used acceptance rules is the Metropolis-Hastings acceptance probability:
\begin{equation}
\label{eq:acceptanceMC}
A(s|s_0) = \min \left(1, \frac{\pi(s) Q(s_0 | s)}{\pi(s_0) Q(s | s_0)}.\right)
\end{equation}
Since $\pi(s)$ and $\pi(s_0)$ appear in the formula as a ratio, the intractable normalization constant $\mathcal{Z}$ cancels out, reducing the computation to evaluating $e^{-\beta\Delta H_c}$.

The performance of MCMC is evaluated by its convergence rate to the stationary distribution $\pi$. Fast convergence is essential for practical applications. The convergence rate is defined by the mixing time $t_{\text{mix}}$, which is the number of steps required for the chain to become $\epsilon$-close (in total variation distance) to the stationary distribution. The distribution at any given time step is determined by the repeated application of the transition matrix $P$. The stationary distribution corresponds to the eigenvalue $\lambda_1 = 1$ of the transition matrix, and the spectral gap determines the convergence rate~\cite{levin2017markov}
\begin{equation}
\label{eq:gapdelta}
\delta = 1 - \max_{i \neq 1} |\lambda_i|,
\end{equation}
where $1=\lambda_1 \geq \lambda_2 \geq \lambda_3 \geq \cdots \geq \lambda_d$ are the eigenvalues of $P$. The mixing time can be bounded by the spectral gap as follows:
\begin{equation}
    (\delta^{-1} - 1) \ln \left(\frac{1}{2\epsilon}\right) \leq t_{\text{mix}} \leq \delta^{-1} \ln \left(\frac{1}{\epsilon \hspace{3pt} \mathrm{min_{s}} \pi(s)}\right).
\end{equation}
A Markov chain with $\delta = 0$ converges immediately, whereas a chain with $\delta = 1$ does not converge at all.

It is important to note that although the spectral gap is an unambiguous metric to compare different MCMC methods, calculating $\delta$ requires diagonalizing the Hamiltonian $H$ to compute the proposal matrix $Q$ and the dense transition matrix $P$. The computational complexity associated with diagonalizing $P$ is  $O\left(2^{3 n}\right)$. Therefore, our analysis is limited to small systems $n<10$. These sizes are consistent with the ones originally investigated in Ref.~\cite{layden2023quantum}.
Note that storing the full transfer matrix $P$ is obviously not necessary in practical Monte Carlo simulations. However, this remains the most accurate method for computing mixing times. To maintain full consistency with the original work in~\cite{layden2023quantum}, we also adopt this strategy. In other words, in this numerical setup, we choose to calculate the mixing times accurately by constructing and diagonalizing an exponentially large, dense matrix.
However, in practical Metropolis simulations, whether quantum or quantum-inspired, this will not be necessary.

\subsection{Quantum-enhanced Markov chain Monte Carlo}
The essence of the quantum-enhanced MCMC algorithm~\cite{layden2023quantum} is to propose new configurations through a quantum step. Specifically, starting from an initial configuration $s_0$, the state is encoded in the computational basis state $\ket{s_0}$. This state is then evolved unitarily under some unitary $U$ and subsequently measured in the computational basis, resulting in a new classical configuration $s$. The proposal probability $Q(s|s_0)$ is determined by the Born rule:
\begin{equation}
    Q(s|s_0) = |\bra{s}U\ket{s_0}|^2.
\end{equation}
The acceptance step remains identical to that in the classical MCMC. However, significant simplification is achieved by selecting a symmetric operator $U$. When $U = U^T$, the Metropolis-Hastings acceptance rule reduces to
\begin{equation}
    A(s|s_0) = \min \left(1, e^{-\beta\Delta H_c} \right),
\end{equation}
because $Q(s_0 | s) = |\bra{s}U\ket{s_0}|^2 = |\bra{s_0}U^T\ket{s}|^2 = Q(s | s_0)$.

The original work considers the evolution under a time-independent quantum Hamiltonian of the spin glass in a transverse field, often referred to as the quantum Sherrington-Kirkpatrick model or quantum spin glass~\cite{yamamoto1987perturbation, young2017stability, guo1994quantum}. The Hamiltonian is given by
\begin{equation}
    H = (1-\gamma) \alpha H_c + \gamma H_\textrm{mix}
    \label{eq:time_independent_hamiltonian}
\end{equation}
where $H_\textrm{mix}$ is a mixing Hamiltonian with non-zero off-diagonal elements to generate transitions. Specifically, $H_\textrm{mix}$ is defined as
\begin{equation}
    H_\textrm{mix} = \sum_{i=1}^{n} X_i.
\end{equation}
The scale factor $\alpha = \frac{||H_\textrm{mix}||_F}{||H_c||_F}$ establishes a common energy scale between the two components of the Hamiltonian, where $\|\cdots\|_F$ is the Frobenius norm. Finally, the parameter $\gamma$ determines the relative weight of the mixing Hamiltonian. Consequently, the time-evolution operator is given by 
\begin{equation}
\label{eq:eiht}
    U = e^{-iHt},
\end{equation} where $t$ represents the total evolution time. Since both terms in the Hamiltonian in Eq.~\ref{eq:time_independent_hamiltonian} are symmetric, the time-evolution operator $U$ is also symmetric. Practically, $U$ can be approximated using a second-order Suzuki-Trotter expansion~\cite{layden2023quantum}.

In Ref.~\cite{layden2023quantum} the parameters $\gamma$ and $t$ are selected uniformly at random for each proposal step. This heuristic approach eliminates the need for optimization and is based on the idea that even if $\gamma$ and $t$ are occasionally chosen suboptimally, the algorithm will still perform adequately as long as optimal values are selected frequently enough.

Finally, we note that the choice of the mixing Hamiltonian is arbitrary, as long as $U$ remains symmetric and the term allows for ergodic exploration of the configurational space. This holds true for the simple one-body transverse field. In Ref.~\cite{mazzola2021sampling} the idea is introduced in the continuum, where an unambiguous choice for the mixing (or kinetic) term is possible: the Fokker-Planck Hamiltonian, which fully determines the equilibrium and kinetic properties of a system.
In principle, as we will see in Sec.~\ref{ss:qaoa}, one could go beyond this structure and employ unitaries which are not strictly defined by Hamiltonian simulation~\cite{nakano2024markov}.

\begin{figure*}[!t]
    \centering
    \includegraphics[width=0.9\textwidth]{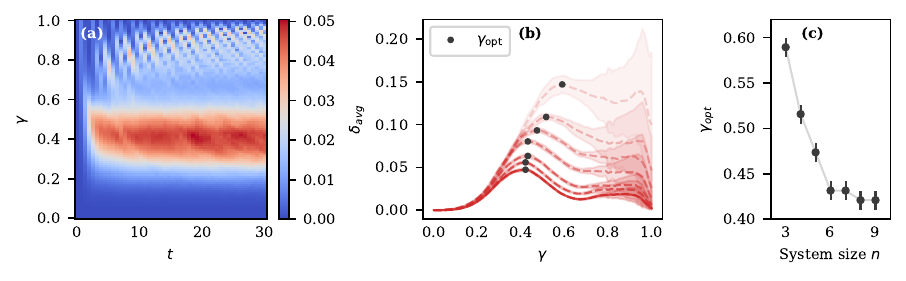}
    \caption{(a) Grid search over $\gamma$ and $t$ for the spectral gap $\delta$ for the $n=9$ system. Throughout this study, each data point is an average over 100 model instances and $T=1$. Warmer colors indicate higher $\delta$, i.e., faster MCMC convergence. (b) Spectral gap $\delta_{\text{avg}}$ averaged over $t \in [12,30]$ as a function of $\gamma$. The red dashed lines represent systems $n=4-8$, and the red solid line represents $n=9$. (c) Optimal gamma values $\gamma_{\textrm{opt}}$ plotted as a function of system size. Error bars represent the standard deviation.}
    \label{fig:grid_search_panel}
\end{figure*}

\subsection{Thresholds for practical quantum advantage}
\label{s:Thresholds for practical quantum advantage}

The main result of Ref.~\cite{layden2023quantum} is that QEMC is able to achieve a scaling advantage for the mixing time compared to all possible classical, local, and cluster Monte Carlo update schemes. In particular, the dependence of the gap 
$\delta$ [Eq.~(\ref{eq:gapdelta})] on the system size $n$ is analyzed. The gap closes exponentially (meaning that the mixing time increases exponentially), as $\delta \propto 2^{-k n}$, with $k>0$, for all schemes considered. However, the QEMC exponent is 
$k_{q}=0.264(4)$, while the classical sampler shows $k_c = 0.94(4)$, resulting in a polynomial speed-up factor of 
$\alpha= k_c/ k_q = 3.6(1)$ (see also Fig.~\ref{fig:scale_factors}).
This means that QEMC features smaller autocorrelation times, allowing it to achieve a target sampling quality with fewer samples compared to classical MCMC, which has a longer mixing time.

Unfortunately, the theoretical scaling advantage can still be overshadowed by the large gate-time prefactor, even for reasonably large system sizes~\cite{babbush:2021}. Note that the time required to execute a quantum gate is several orders of magnitude slower than that of a classical one. Accepted estimates for the gate clock of future fault-tolerant hardware are on the order of 10 kHz~\cite{gidney2019efficient}. This implies that the quantum algorithm starts at a significant disadvantage compared to a classical machine, which is not compensated by more favorable scaling until a certain crossover problem size is reached.
Current noisy machines operate at higher frequencies but lack error correction.
More precisely, the classical and quantum runtimes are $\mathcal{T}_c (n) \propto t_c(n)~ 2^{k_c n}$ and $\mathcal{T}_q (n) \propto t_q(n) ~2^{k_q n} $, respectively, where $t_{c/q}(n)$ are the runtimes to perform one single MCMC step in both schemes.
This includes not only the algorithmic scaling with the system size but also the hardware gate-time prefactor. The crossover size at which the quantum algorithm outperforms the classical one is at
\begin{equation}
    n_{\textrm{threshold}} > \frac{1}{k_c - k_q} \log_2 \left( \frac{t_q(n)}{t_c(n)}\right).
\end{equation}
Therefore, the ratio of the quantum and classical runtimes per iteration $t_q/t_c$ is logarithmically suppressed compared to the inverse difference between the scaling exponents.

We attempt a rough, order-of-magnitude estimate of the conditions that must be met for a practical advantage on a concrete case study. We follow the discussion in Ref.~\cite{lemieux2020efficient}, where quantum walks are studied.
First, it is necessary to identify the classical state of the art and a problem size that is challenging for current methods. In Ref.~\cite{lemieux2020efficient} the Janus field-programmable gate array–based special-purpose machine is selected as a benchmark. This hardware was built to simulate a $80  \times 80 \times 80$ cubic spin glass with local couplings and can perform about one spin flip per picosecond. The numerical experiments involve a total of $10^{18}$ local updates (spin flip) and run for about one month~\cite{Janus2012}.

These parameters establish a threshold for quantum advantage under different scenarios. Assuming that QEMC has a true scaling advantage of order $\alpha$, this means it could achieve the same sampling quality, i.e., accuracy on computing thermodynamic averages, with $10^{18/\alpha}$ samples.
For the sake of concreteness, let us assume values of $\alpha = 4, 3, 2$. For $\alpha = 4$, i.e. quartic speedup, we find that the crossover for quantum advantage occurs at outputting $10^{4.5}$ samples in one month, which means performing about one quantum proposal step per minute.
Conversely, if $\alpha = 3$, it would require performing $10^6$ QEMC steps in one month, meaning one step every 2.5 s.
The observed speed up in Ref.~\cite{layden2023quantum}, 3.6(1), stands in between these values.
Finally, if the speed up is only quadratic, the possibility of practical quantum advantage diminishes, as it would require performing one QEMC step every 2 ms.

While these constraints seem within reach, the quantum circuit in this case is defined over about 500,000 qubits.
In this example, the system's size is very large, but the couplings are local. 
Fully connected models become hard at much lower sizes~\cite{barahona1982computational}

Another critical consideration is the scaling of the Hamiltonian simulation time $t$ in Eq.~(\ref{eq:eiht}), and consequently the circuit depth, with the system size, $n$. Obviously, a scaling dependence $t\propto 2^{n\beta}$ will decrease the order of quantum speedup from $\alpha = k_c / k_q$ to $\alpha = k_c / (k_q + \beta)$. However, in Sec.\ref{s:t} we show that the total evolution time likely scales subexponentially and thus does not affect asymptotic quantum speedup.

\section{Optimal parameter regimes}
\label{s:optparam}
The first objective of the paper is to gain a deeper understanding of the parameter regimes in which the quantum proposal strategy is effective. We perform the analysis without actual quantum hardware by computing the spectral gap numerically using exact continuous-time limit integration of the quantum dynamics across a wide range of $\gamma$ and $t$ values for varying system sizes. Fig.\ref{fig:grid_search_panel} (panel \textit{a}) displays the landscape of $\delta$ values for a $n=9$ qubit system. A similar grid search for smaller systems reveals consistent results. 

In general, the relationship between $\delta$ and the parameters $\gamma$ and $t$ in Figure\ref{fig:grid_search_panel} (a) appears complex, but some intuition can be gained by examining the limiting cases, similarly to Ref.~\cite{orfi2023near}. When $\gamma=1$, the Hamiltonian consists solely of the mixing term, leading to a single-qubit rotation around the x axis:
\begin{equation}
    U = e^{-iH_\textrm{mix}t} = \prod_{i=1}^{n} e^{-iX_it}.
\end{equation}
This produces oscillatory behavior along the $t$ axis. When measured in the computational basis, this quantum evolution works as a classical proposal, flipping each spin with a probability of $\sin^2(t)$. For example, at $t = \frac{\pi}{4}$, the process reduces to the random proposal strategy. On the other hand, for sufficiently small $\gamma$, the landscape can be understood using perturbation theory~\cite{layden2023quantum}. The evolution under the classical Hamiltonian $H_c$ perturbed with a small mixing term $H_\textrm{mix}$ generates transitions between computational eigenstates $\ket{j}$ and $\ket{k}$ with a probability given by:
\begin{equation}
    P(\ket{j} \rightarrow \ket{k}) \propto |\bra{k}H_\textrm{mix}\ket{j}|^2\gamma^2 + O(\gamma^3).
\end{equation}
Since $H_{\textrm{mix}}$ is a sum of $X_i$ terms, this evolution only induces transitions between configurations that differ by a single spin flip. This effectively reduces the process to the classical local proposal strategy.

\subsection{Optimal \texorpdfstring{$\gamma$}{gamma}}
Remarkably, there is a pronounced region of high spectral gap for $\gamma$ values in the range $[0.3, 0.5]$. Averaging over $t$ in the range $[12,30]$, we observe peaks in $\delta$ across all system sizes, as illustrated in Fig.~\ref{fig:grid_search_panel} (b). The corresponding optimal $\gamma_{\textrm{opt}}$ values are plotted for each system size in Fig.~\ref{fig:grid_search_panel} (c). We see that $\gamma_{\textrm{opt}}$ decreases monotonically with system size $n$, flattening out around $\gamma=0.42$ for the $n=9$ system. The apparent convergence to an asymptotic value is physically intuitive, although verification would require larger-scale simulations. If an asymptotically optimal value indeed exists, the quantum proposal strategy could be simplified and enhanced by using a fixed optimal $\gamma$ value rather than picking $\gamma$ at random.

We observe that values of $\gamma$ leading to the highest spectral gaps lie in the region close to the expected phase transition of the quantum-spin-glass model~\cite{young2017stability, mukherjee2015classical, guo1994quantum}. To validate this observation, we compute the phase diagram with respect to $\gamma$ and $T$ in Appendix~\ref{s:pt} and determine the critical gamma value $\gamma_c=0.50\pm0.02$. This value is close to but larger than the optimal $\gamma_{\textrm{opt}}$ determined by grid search. Therefore, we conjecture that critical behavior near the phase transition, characterized by enhanced quantum fluctuations, helps to generate a good proposal distribution. However, the exact mechanism underlying this relationship remains unclear and needs further investigation.

\subsection{Total evolution time}
\label{s:t}
\begin{figure}[!tb]
    \centering
    \includegraphics{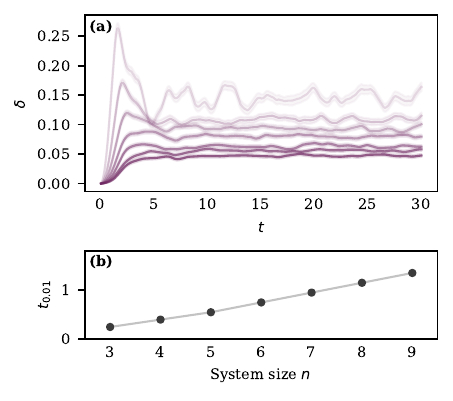}
    \caption{(a) Spectral gap as a function of the total evolution time $t$. Systems of sizes $n=3-9$ are shown. (b) Time $t_{\mathrm{0.01}}$ to hit a fixed gap value $\delta=0.01$ as a function of system size $n$.}
    \label{fig:delta_vs_t}
\end{figure}

To understand the role of the total evolution time, we fix $\gamma$ at its optimal value for each size of the system and compute the absolute spectral gap as a function of total evolution time. As shown in Fig.~\ref{fig:delta_vs_t} (a), the gap increases rapidly at short times before stabilizing and fluctuating around a steady value at longer times. This suggests that, beyond a certain point, the gap becomes only weakly time dependent.

A key question is how the time to reach this steady value scales with system size. Due to significant fluctuations, precise measurement of the convergence time is challenging. Instead, we calculate the first time the gap crosses a fixed value, $\delta = 0.01$, for all system sizes. The results, shown in Fig.~\ref{fig:delta_vs_t} (b), indicate linear scaling, suggesting that the quantum speedup remains unaffected asymptotically (see Sec.~\ref{s:Thresholds for practical quantum advantage})

This also highlights that the original strategy of drawing $t$ randomly from a fixed range $[2, 20]$ may not be viable for large systems, as the appropriate time range likely needs to scale with the system size.

\begin{figure}[!tb]
    \centering
    \includegraphics{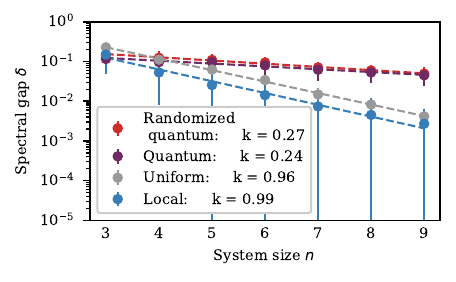}
    \caption{Spectral gap $\delta$ for the four different proposal strategies. Randomized quantum refers to the original strategy from Ref.~\cite{layden2023quantum}. Quantum refers to the quantum strategy with fixed parameters $\gamma=0.45$ and $t=12$. Exponential fits are shown as dotted lines with the corresponding fit parameters. Error bars represent the standard deviation.}
    \label{fig:scale_factors}
\end{figure}

Empirically, we find that setting $\gamma$ to a fixed value $\gamma=0.45$, slightly below its critical value, and setting fixed at $t=12$ slightly outperforms the original fully randomized strategy, as shown in Fig. \ref{fig:scale_factors}. 
While these observations cannot be considered conclusive as they are based on a fairly limited number of spins, we do not identify any clear exponential bottleneck that may have been previously overlooked.
Overall, it is interesting to note that the randomized strategy is already competitive with respect to a more fine-tuned parameter choice.

\section{Optimized schedules and QAOA}
\label{s:optsched}

\subsection{Time-dependent Hamiltonian evolution}

The surprisingly good performance of the original randomized strategy, challenges us to explore different types of circuits.
The quantum proposal method is not limited to time-independent Hamiltonians. Time-dependent Hamiltonians can also be employed for time evolution, provided they satisfy the symmetry constraint $U(\tau) = U(\tau)^T$, where $\tau$ represents the total evolution time. Reference ~\cite{layden2023quantum} demonstrates that a sufficient condition is $H(t) = H(t)^T$ and $H(t) = H(\tau - t)$ for all $t \in [ 0, \tau]$. In the following, we introduce the time-dependence by allowing $\gamma(t)$ to vary with time:
\begin{equation}
    H(t) = [1-\gamma(t)] \alpha H_c + \gamma(t) H_\textrm{mix}
    \label{eq:time_dependent_hamiltonian}
\end{equation}
To maintain symmetry, $\gamma(t)$ must be symmetric about the midpoint of the time evolution. We note that \textit{a priori}, it is unclear whether evolving under a time-dependent Hamiltonian would be beneficial for proposing MCMC moves. Moreover, even if there is potential for enhancement, the optimal form of $\gamma(t)$ is unknown.

Therefore, in this study, we adopt an uninformed hands-off approach by using a classical optimizer to design the optimal $\gamma(t)$ schedule, inspired by recent variational quantum adiabatic algorithms for ground state preparation~\cite{schiffer2022adiabatic, matsuura2021variationally, passarelli2022optimal, finvzgar2024designing}. These algorithms use classical optimization to adjust adiabatic evolution parameters to the Hamiltonian gap structure, for example, by changing the annealing speed to minimize Landau-Zener transitions when the evolution passes through the region with the minimal spectral gap.

However, we are addressing a fundamentally different problem here. Landau-Zener transitions, which are typically detrimental in the context of adiabatic quantum evolution, can be a valuable resource in our case if they occur between states that are close in energy but potentially far in Hamming distance~\cite{layden2023quantum}. Thus, it is worth investigating whether a classical optimizer can identify and leverage such diabatic effects by adjusting the time-evolution shape of our Hamiltonian~\cite{crosson2021prospects}. In this context, we consider employing Bayesian optimization (BO) to design an optimized schedule for $\gamma(t)$. Bayesian optimization is a widely used global optimization technique~\cite{mockus2005bayesian}, particularly effective when the cost function is expensive to evaluate, such as in computing the absolute spectral gap. We refer the reader to Appendix~\ref{sec:bayesian_optimization} for a brief overview of BO and more details.

\begin{figure}[!t]
    \centering
    \includegraphics{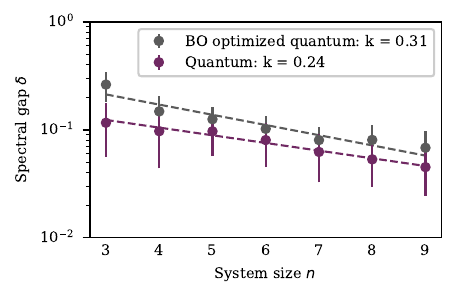}
    \caption{Absolute spectral gap $\delta$ for the fixed $t = 12$ and $\gamma = 0.45$ quantum proposal (purple) and the quantum proposal based on the classically optimized time-dependent Hamiltonian (dark grey). Exponential fits are shown as dotted lines with the corresponding fit parameters.}
    \label{fig:scale_factors_ra}
\end{figure}

The spectral gaps achieved with these optimized schedules are presented in Fig.~\ref{fig:scale_factors_ra}. Bayesian-optimization-optimized schedules outperform the time-independent strategy with fixed $t$ and $\gamma$ across all system sizes. However, the performance gains are irregular and the resulting scaling coefficient is in fact worse. For $n \in [8,9]$, BO-optimized schedules closely mimic the time-independent strategy. A key distinction is that the optimized schedules ramp up the transverse field gradually, whereas the time-independent approach applies it instantaneously.

Our findings indicate that the classical optimizer can indeed surpass the original strategy for smaller systems, potentially by leveraging strong diabatic effects. However, for larger systems, the optimizer tends to converge to a time evolution similar to that of the original strategy. Overall, the fact that this method performs irregularly leads to worse scaling.

\subsection{Symmetric QAOA circuit}
\label{ss:qaoa}

An alternative way of constructing the quantum proposal is utilizing a parametrized, symmetric ansatz.
Reference ~\cite{nakano2024markov} proposes a variational circuit inspired by the quantum approximate optimization algorithm (QAOA) and tests it against classical update.
However, the extension is not tested against the original QEMC proposal, so we need to understand if this choice actually constitutes an improvement over the original method.

\begin{figure*}[!tb]
    \centering
    \includegraphics[width=0.9\textwidth]{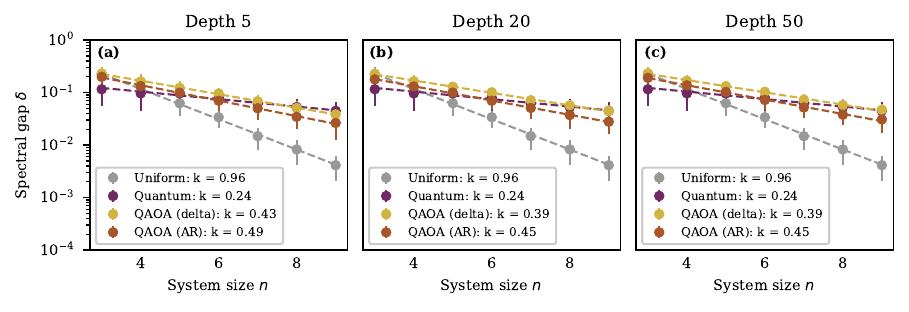}
    \caption{QAOA-inspired circuit. The variation of the spectral gap with system size $n$ is plotted for circuit depths (a) $p = 5$, (b) $p = 20$ and (c) $p = 50$. An increase in circuit depth from $p = 5$ to $20$ results in a small improvement in the spectral gap, but no further improvement is observed when increasing the depth to $p = 50$. Each plot represents data obtained from 100 random instances. The parameters used for the quantum strategy are $\gamma = 0.45$, $t = 12$, and $T = 1$.}
    \label{fig:delta_vs_n_qaoa}
\end{figure*}

The ansatz has the form
\begin{equation}
    U=V(\boldsymbol{\beta}, \boldsymbol{\gamma})^{\top} V(\boldsymbol{\beta}, \boldsymbol{\gamma}),
\end{equation}
where
\begin{equation}
    V(\boldsymbol{\beta}, \boldsymbol{\gamma})=U_C\left(\gamma_p\right) U_B\left(\beta_p\right) \cdots U_C\left(\gamma_1\right) U_B\left(\beta_1\right),
\end{equation}
with 
$ U_B(\beta)= e^{-i H_{\textrm{mix}} \beta}$ and $U_C(\gamma)=e^{ -i \alpha H_{c} \gamma}$.
It features $p$ layers, implying that there are $2p$ parameters to optimize: $\beta = \{\beta_1, \cdots, \beta_p\}$ and $\gamma = \{\gamma_1, \cdots, \gamma_p\}$.
This circuit resembles the structure of a discrete reverse-annealing process, which has also been proposed in Reference ~\cite{layden2023quantum} as a possible quantum process for the method.
However, this circuit can be more general and, most importantly, can offer a constant-depth ansatz.

Following Ref.~\cite{nakano2024markov}, we simplify the optimization process, using the constraint $
\theta = \beta_1 = \cdots = \beta_p = \gamma_1 = \cdots = \gamma_p.
$.  
This is necessary as optimizing from scratch QAOA parameters without any prior guess~\cite{zhou2020quantum} has proven to be inefficient, especially when the cost function comes with a statistical error bar~\cite{scriva2024challenges}.

Unlike the plain QAOA, where the cost function is an energy, the objective here is to minimize the mixing time of the MCMC. The practical approach proposed in Ref.~\cite{nakano2024markov} is to perform short MCMC runs and compute the acceptance rate. While this argument is not rigorous, the authors of Ref.~\cite{nakano2024markov} found a correlation between $\delta$ and the acceptance rate, which holds only for small values of $\theta$. Therefore, the practical strategy would consist of minimizing the acceptance rate (AR) under these conditions.
In turn, the AR can be measured empirically from MCMC runs, or exactly from the definition of the $Q$ and $A$ matrices, though this would be exponentially costly.

However, we expect that directly maximizing $\delta$ will lead to the best result the algorithm can offer, as opposed to minimizing the acceptance rate. Since our goal is to assess whether the entire QAOA ansatz is beneficial at all, we also choose to try this second objective. By directly maximizing $\delta$, as in the previous sections, we aim to test the performance of the variational circuit without further assumptions. While also not scalable, this approach provides a more straightforward benchmark for the small system sizes considered here.
In this section, we perform Statevector simulation of the circuits using \textsc{qiskit} software package~\cite{qiskit2024}. We plot the results in Fig.\ref{fig:delta_vs_n_qaoa}, showcasing the performance for various depths \( p \). While there is a slight improvement in the gap between \( p=5 \) and \( 20 \), it is evident that, in this case as well, the scaling of the gap versus system size is not improved compared to the original approach.

\section{quantum inspired algorithms}
\label{s:tns}

In the previous sections, we observed the remarkable empirical superiority of the original Hamiltonian dynamics approach. This section explores whether this protocol is robust against algorithmic and approximation errors.
The first point allows us to determine whether we truly need to perform the quantum dynamics simulation in its continuous-time limit, thus saving resources.
The second point could establish a new class of classical update schemes that are only inspired by the QEMC protocol but do not require quantum hardware.

\subsection{Trotter error}
\label{ss:trotter}

\begin{figure*}[!tb]
    \centering
    \includegraphics[width=0.9\textwidth]{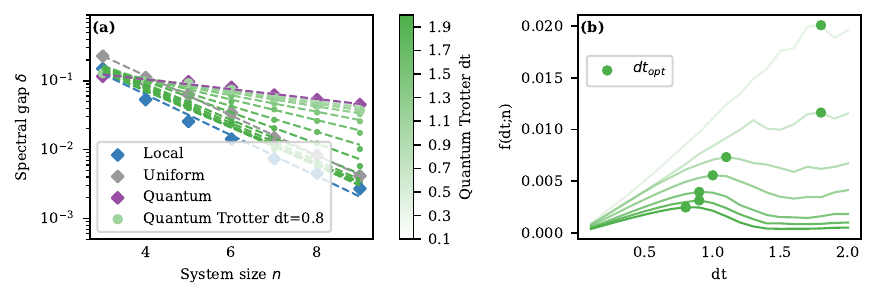}
    \caption{(a) Impact of Trotter step size $dt$ on the quantum proposal strategy. (b) Objective function to find optimal Trotter step size $dt_{\text{opt}}$ for different system sizes. The optimal step size converges to $dt=0.8$ for $n=9$. Here the quantum and quantum Trotter strategies use $\gamma=0.45$ and $t=12$.}
    \label{fig:spectral_gap_optimal_trotter}
\end{figure*}

To simulate the real-time dynamics, we use a second-order Trotter scheme, which, for this Hamiltonian,  essentially has the same computational cost as the first-order Trotter formula.
This is because the Hamiltonian consists of only two non-commuting operators ~\cite{layden2023quantum}. The first-order and second-order decompositions differ only by a phase, which is irrelevant when the wave function is measured.
First of all, in Appendix~\ref{app:detailbalance}, we show that any Trotter error preserves the symmetry of the proposal. However, the efficiency of the algorithm may vary.

In the original paper, a Trotter time step of 0.8 was chosen, primarily to meet hardware constraints ~\cite{layden2023quantum}. Indeed, it has been shown, both theoretically~\cite{knee2015optimal} and experimentally ~\cite{layden2023quantum,miessen2024benchmarking}, that a finite time step is optimal in the presence of hardware gate noise. Moreover, a large time step allowed for the practical implementation of the $e^{-i \theta/2 ~ Z \otimes Z}$ gate, using pulse-efficient cross-resonance gates ~\cite{earnest2021pulse}. 

Here we run the QEMC algorithm using $t=12$ and $\gamma=0.45$, using discrete-time evolution methods. We vary the Trotter step $dt$ and analyze the gap scaling.
The results are reported in Fig.~\ref{fig:spectral_gap_optimal_trotter} (a).
We observe that different values of $dt$ lead to a spectrum of results. Large values of $dt$ yield a scaling similar to uniform inefficient proposals, while we recover the continuous-time result only when $dt \rightarrow 0$. Interestingly, finite $dt$ never produces better scaling than the continuous-time result. 

There is however a subtlety here. While smaller values of $dt$ improve the scaling, they require more Trotter steps $m$ to reach a target $t = m~dt$. This remains true even when $t$ is sampled randomly. To identify 
the computationally optimal $dt$, we maximize the objective function
\begin{equation}
\label{eq:target_dt}
    f (dt;n) = \frac{\delta(dt;n) ~ dt} { t},
\end{equation}
where $t=12$ is a constant. 
Equation~\ref{eq:target_dt} follows from the definition of $\delta$ as the inverse mixing time and the computational cost growing with decreasing $dt$.
We observe that this objective function shows indeed a maximum, whose position depends on $n$.
Results in Fig.~\ref{fig:spectral_gap_optimal_trotter}(b) suggest that the optimal time steps concentrate, for the larger $n$, around the value of $dt_{\text{opt}}=0.8$. 
Surprisingly, this is the same value chosen for the hardware simulation of Ref.~\cite{layden2023quantum}, though for different motivations. We receive the same value for $dt_{\text{opt}}$ when running the QEMC algorithm in its original setting, i.e., randomizing over the total evolution time $t$ and $\gamma$.
The simulations in these sections were done using the \textsc{Julia} programming language \cite{Julia-2017} and using 48 CPUs with 2560 MB memory each of the ETH Euler Cluster. Individual instances were calculated in parallel using multi-processing and the \textsc{Distributed.jl} library .

\subsection{A tensor network quantum inspired version}
\label{ss:mps}

The novelty of the QEMC algorithm lies in using a quantum computer for the proposal step in MCMC. The Hamiltonian time evolution can be performed using a Trotterized approach for digital platforms or could possibly be implemented directly on analog simulators.
A question arises: Do we really need a quantum computer for this?
While it is true that Hamiltonian evolution cannot be exactly simulated classically, the algorithm may not require exact evolution. It is important to note that, given the existence of the acceptance step, approximation errors will not propagate into sampling errors. Their only effect could be on performance, similar to the impact of hardware noise~\cite{layden2023quantum}.
Thus, we ask whether a classical approximation of the process is detrimental or not.

This exploration is not intended to disprove the (so far empirical) quantum advantage of QEMC, but rather to provide a viable near-term alternative while we wait for the next generation of quantum hardware needed to run the algorithm at scale.

\begin{figure*}[!tb]
    \centering
    \includegraphics[width=0.9\textwidth]{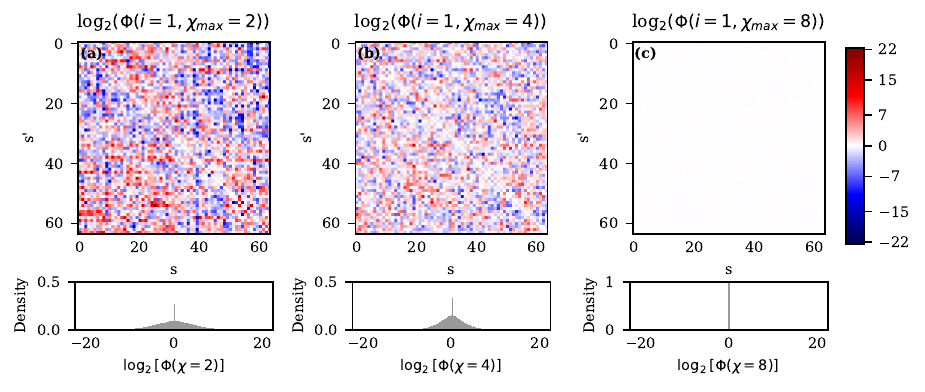}
    \caption{Impact of different bond dimensions $\chi$ on the fraction $\Phi(s,s',i,\chi)$ for $n=6$, $t=12$, $\gamma=0.45$, $dt=0.8$, and (a) $\chi = 2$, (b) $\chi = 4$ and (c) $\chi = 8$. The fraction is rescaled to $\log_2(\Phi)$. The heatmap represents $\Phi(s,s',\chi)$ for one instance. The density is the histogram for the rescaled fractions $\log_2[\Phi(s,s',i,\chi)]$ given for 100 instances. $\log_{2}(\Phi=1)=0$ indicates that the detailed balance is not violated. We observe large deviations for $\chi < 2^{n/2}$.
    }
    \label{fig:DetailedBalancePlot.pdf}
\end{figure*}

Here we choose to adopt matrix product state (MPS) simulations~\cite{schollwock2011density, orus2014practical}. However, other options are possible, ranging from other types of tensor networks~\cite{orus2014practical, verstraete2004renormalization, murg2010simulating} and neural networks ~\cite{donatella2023dynamics,sinibaldi2023unbiasing} to time-dependent variational Monte Carlo~\cite{carleo2012localization,sinibaldi2023unbiasing}.
The key point we aim to investigate is whether the classical approximation error qualitatively prevents us from running the resulting quantum-inspired algorithm.
In MPSs or generic tensor-network states, the bond dimension $\chi$ controls the approximation level of the true quantum state.
While the exact wave function of a system of $n$ spin needs  $2^n$ complex-valued coefficients, MPSs allow us to compress the wave function from $2^n$ to $O(2 n \chi^2)$ coefficients. Any wave function can be exactly represented by an MPS with a bond dimension of $\chi = 2^{n/2}$, but for practical computations, $\chi$ needs to be fixed at a much smaller threshold.

We perform the evolution using the time-evolving block decimation (TEBD) method~\cite{vidal2004efficient,paeckel2019time}.
When performing a time evolution, each application of a two-site gate on the MPS doubles the bond dimension of the associated bond indices if there is no low-rank approximation and the bond dimension is not truncated. At each step, the bond dimension is restored at the price of a truncation error.
The longer the time evolution is, the larger the approximation error will be, at fixed $\chi$. We use the software package \textsc{itensor.jl}~\cite{fishman2022itensor}.

\subsubsection{Symmetry error}

The first thing to check is whether the approximation error disrupts the symmetry of the proposal.
We define the ratio
\begin{equation}
    \Phi(s,s',n, t, \chi) = \frac{Q(s|s'; n, t, \chi)}{Q(s'|s; n, t, \chi)}, 
\end{equation}
which quantifies the impact of the bond dimension on the symmetry condition. In principle, $Q(s|s'; n, t, \chi) \in [0,1]$, which implies that $\Phi(s,s',n, t, \chi) \in [0, \infty)$; $\Phi(s,s',n, t, \chi)=1$ indicates that the symmetry conditions is fulfilled.

Figure ~\ref{fig:DetailedBalancePlot.pdf} displays the impact of the bond dimension on the fraction $\Phi$ for $n=6$, $t=12$ and $dt=0.8$ for different $\chi \in \{2,4,8\}$. The figure shows that $\log_{2}(\Phi)$ is distributed around $0$ and gets more peaked for an increasing $\chi$. In the limit of $\chi=2^{n/2}$ the $\log_{2}(\Phi)$ resembles a $\delta$ function, demonstrating that we need a bond dimension of $\chi = 2^{n/2}$ to get $\Phi=1$ with float point precision. Hence, we have to incur a large bond dimension $\chi = 2^{n/2}$ to be able to neglect the fraction.

\begin{figure}[!tb]
    \centering
    \includegraphics{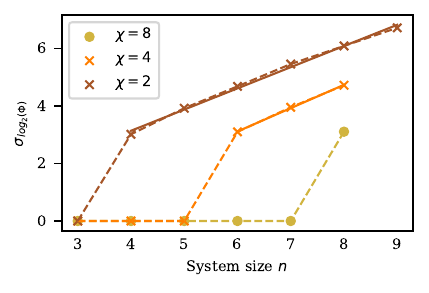}
    \caption{Standard deviation of the logarithmically rescaled fraction $\log_2(\Phi)$ averaged over 100 instances for MPS proposal strategies with different bond dimensions $\chi$ for $\gamma=0.45$ and $t=12$.}   \label{fig:detailed_balance_std_logfraction_graphfull_n_03_09_gammaRandom_dt_01_01_20_samples001to096_chi_2_4_8.pdf}
\end{figure}

We investigate the scaling of $\Phi$ with the system size $n$ for different $\chi$. In this case, it is again convenient to plot the logarithm of this quantity,
because if $\Phi=1$, $\log_2 \Phi=0$. To quantify the symmetry error due to the bond dimension, we look at the standard deviation $\sigma_{(\log_2 \Phi)}$. Figure ~\ref{fig:detailed_balance_std_logfraction_graphfull_n_03_09_gammaRandom_dt_01_01_20_samples001to096_chi_2_4_8.pdf} displays this quantity averaged over 100 different instances for different bond dimensions. We observe that this grows linearly in $n$, which implies a much larger deviations for $\Phi$.% an exponential growth for $\Phi$. 

Summarizing the results of this section, we find that the symmetry condition is violated for a bond dimension smaller than the one required for an exact representation of the state, i.e. $Q(s'|s) \neq Q(s'|s)$ for $\chi < 2^{n/2}$. As expected, we find that the error on the fraction grows in system size for a fixed bond dimension.

This however is not a no-go type of result, as it only implies an extra computation for the acceptance step. We need to calculate $Q(s|s'; n, t, \chi)$ and its reverse $Q(s'|s; n, t, \chi)$ for each MCMC step and include the ratio $\Phi = Q(s|s')/Q(s'|s)$ explicitly in the acceptance step [see Eq.~(\ref{eq:acceptanceMC})]. While the symmetry requirement is crucial for the quantum implementation of QEMC, it only doubles the computation in the classical case, if the ratio $\Phi$ can be evaluated efficiently in a numerically stable way.

\subsubsection{Scaling of the approximate MPS-QEMC}
\label{subs:Scaling of the approximate MPS-QEMC}

We now analyze whether this quantum-inspired strategy has the potential to be advantageous compared to traditional classical updates. To do so, we perform the usual gap scaling analysis. We apply Trotter evolution with the MPS, using three different choices of bond dimensions. For simplicity, we use the optimal time step identified previously. Although we already confirmed that a bond dimension $\chi$ cannot accurately represent the exact time-evolved state for $\chi < 2^{n/2}$, it may still be useful for the proposal step.

\begin{figure}[!tb]
    \centering
    \includegraphics{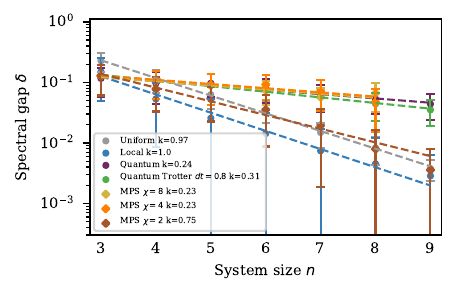}
    \caption{Spectral gap $\delta$ for different system sizes $n$ at fixed temperature $T=1.0$ for different proposal strategies, with$\gamma=0.45$ and $t=12$.  Trotter $dt = 0.8$ refers to the Trotterized quantum proposal. MPS $\chi$ refers to using TEBD with the MPS to perform the time evolution with bond dimension $\chi$. Each data point is averaged over 100 instances. The MPS proposals are able to replicate the quantum proposal with a bond dimension $\chi<2^{n/2}$.}
    \label{fig:MPSgapscaling}
\end{figure}

The results, plotted in Fig.~\ref{fig:MPSgapscaling}, are interesting. We observe that the scaling coefficient $k$ of the quantum-inspired, yet classically simulatable, version of QEMC can vary widely depending on the bond dimension. For instance, for a small bond dimension $\chi=2$ (in brown), the algorithm underperforms and essentially performs like the uniform strategy. If we increase the bond dimension slightly, we observe a scaling similar to or even slightly better than that of the exact and Trotterized quantum proposals. While the system sizes considered are still small, we explore ranges that cannot be exactly captured by the chosen bond dimension.

Specifically, an MPS representation with $\chi = 4$ cannot be exact for $n > 5$, yet we are able to follow the ideal scaling ($k = 0.23$) up to $n = 9.$
This suggests the possibility of achieving, in principle, a scaling advantage through approximate classical emulation of the quantum process.
However, even in this setup, our numerical experiments remain quite expensive, as they require constructing and diagonalizing the full proposal matrix. Further work will be needed to optimize the network architecture and extend the approach to larger system sizes.

\subsubsection{Threshold for practical quantum inspired advantage}
\label{s:Threshold for practical quantum inspired advantage}

In the preceding section, we showed that the quantum-inspired Monte Carlo (QIMC) can in principle replicate the spectral gap scaling of the QEMC using bond dimensions much smaller than what is needed for an exact simulation ($\chi \ll 2^{n/2}$). However, as mentioned in Sec.~\ref{s:Thresholds for practical quantum advantage}, the spectral gap scaling does not account for the runtime to perform a single Markov chain step $t_{qi}(n)$. Similarly to the QEMC, there is a large gate-time prefactor for the QIMC compared to other classical alternatives $t_{qi}(n)\gg t_{u}(n), t_{l}(n)$ if we use uniform ($u$) and local ($l$) as the benchmark. 

Therefore, we attempt an order-of-magnitude estimate of the conditions that must be met for a practical advantage of the QIMC algorithm over the benchmark alternatives given that the QIMC can retain the polynomial speed up. In the following, we look at a best-case scenario, where we assume that the scaling advantage of the QEMC carries over to the QIMC, i.e. $k_{qi} \approx k_{q}$. To compare the runtime per Markov chain step,  we use the computational cost estimates summarized in Table ~\ref{tab:computational_cost} as proxies for the runtime of a single Monte Carlo time step. 

\begin{table}[h]
    \centering
    \begin{tabular}{c c c} 
        Proposal & Memory & Time \\
        \hline
        Local & $O(n)$ & $O(1)$ \\
        \hline
        Uniform & $O(n)$ & $O(n)$ \\
        \hline
        MPS & $O\left(2 n \chi^2+ 16 n^2 \right)$ & $O\left(8 m n^3\chi^3\right)$
    \end{tabular}
    \caption{Memory and computational time estimates for different proposals strategies. The derivation of the quantities is detailed in Appendix~\ref{app:Memory and Computational Cost Estimates of TEBD for MPS}.}
    \label{tab:computational_cost}
\end{table}

We use the computational time $O(1)$ of the local proposal to approximate $t_{c} = 1$ and $t_{q,i} \left(m n^3\chi^3 \right)$ for the TEBD of MPS. This results in

\begin{equation}
    n_{\textrm{threshold}} > \frac{1}{k_c - k_{qi}} \log_2 \left( m n^3\chi^3  \right) 
\label{eq:quantum_inspired_threshold}
\end{equation}

Equation~\ref{eq:quantum_inspired_threshold} indicates that the logarithm suppresses the system size $n$, evolution time $t$ represented by the number of Trotter steps $m$, the bond dimension $\chi$, and a possibly large pre-factor, compared to the inverse difference between the scaling exponents. In addition, Eq.~\ref{eq:quantum_inspired_threshold} highlights the dependence of the algorithm on how the evolution time $t$ (and number of Trotter steps $m$) and the bond dimension $\chi$ scale with system size $n$. 

Clearly, for the exact emulation of the algorithm, $\chi$ would scale exponentially, i.e. $\chi \propto 2^{k_{\chi} n}$. This would decrease the order of the speedup from $\alpha = k_c / k_q$ to $\alpha = k_c / (k_q + k_{\chi})$. 
We know that for an exact representation of the state $k_{\chi}=0.5$. 
Note also that, under these conditions, the algorithm could not be exactly simulated classically due to memory limitations, so its asymptotic scaling should not even be discussed.

The purpose of this section is to highlight the subtlety of achieving a practical advantage with the quantum-inspired version of the method. This approach could be viable if the following conditions are met: (i) the existence of a finite 
$\chi$ such that a gap enhancement at a problem size 
$n$ is still present (compared to conventional classical updates) and (ii) the availability of an efficient classical code where the prefactor does not overshadow this enhancement during runtime.

Note that if a scaling advantage holds, there must necessarily be a threshold system size at which the quantum-inspired proposal outperforms the local update. However, we prefer not to make precise statements about this value, as it is unlikely that the MPS setup presented in this paper is the optimal approach for this task.
These results should be understood as a proof of principle demonstrating the approximate quantum-inspired proposal. The key insight of this study is that exact quantum dynamics may not be required to achieve quantum advantage. This observation aligns with the fact that the quantum hardware implementation in Ref.~\cite{layden2023quantum} is robust against hardware noise.
Ultimately, the best trade-off between approximation and performance will need to be assessed empirically, using larger system sizes and a more efficient method for measuring mixing times, i.e., without requiring the full construction and diagonalization of matrix $P$.

\section{Conclusion}
\label{s:conclu}
We numerically investigated the quantum-enhanced Monte Carlo algorithm\cite{layden2023quantum} and identified its optimal working point.
The core of the algorithm is its quantum-powered proposal step, obtained by time evolving (for a time $t$) an effective Hamiltonian $H$, which is the sum of the classical Hamiltonian $H_c$ and a non diagonal term $H_\textrm{mix}$ multiplied by a scaling factor $\gamma$.
These parameters $\gamma$ and $t$ represent the two major tunable aspects of the algorithm. In the original references, they were chosen at random. Our results show that the optimal $\gamma$ lies close to, but not exactly on, the phase transition of the associated model, while the $t$ appears to scale with the system size.
This introduces only logarithmic corrections to the polynomial asymptotic scaling advantage. 
Additionally, we identify an optimal Trotter step for the real-time evolution.

Recently, Ref.~\cite{orfi2024quantum} suggested that parameter tuning may become more challenging as the system size increases. Since our evidence has been gathered necessarily from small-size systems, we are unable to observe this issue. However, if this turns out to be the case, a simple fix may be again provided by the randomized strategy.
Ultimately, the definitive test of the algorithm’s efficiency will be to run it on large-scale quantum hardware and directly measure its advantage.

We also explored alternative circuit definitions but observed no improvements. The simplest choice appears to be the optimal one.

One of the major novel contributions of this paper is the proposal of a quantum-inspired version of the method. This approach mimics the quantum proposal step using classical hardware. While quantum-inspired classical algorithms have been proposed in the fields of
optimization\cite{han2002quantum,jarret2017substochastic,isakov2016understanding} and machine learning\cite{tang2019quantum,arrazola2020quantum}, in this paper they have been suggested in the context of accelerating classical Markov Chain Monte Carlo methods.

In this work we adopted matrix product states as an approximate emulator of quantum dynamics. The bond dimension serves as a simple hyperparameter that tunes the quality of the approximation. However, other classical strategies could also be employed, from neural networks\cite{sinibaldi2023unbiasing} to tree tensor networks~\cite{shi2006classical}, where two spins are connected through $O(\log(n))$ spins instead of $O(n)$. 
Our results suggest that the original quantum advantage can partially persist even if the emulation is not exact.

Naturally, the runtime per step of the quantum-inspired proposal is orders of magnitude longer than a simple spin flip. Therefore, it is not obvious that this approach will be effective for genuinely large systems. 
For the sake of clarity, the quantum-inspired approach is viable only if all the following assumptions are met:
(i) The asymptotic scaling advantage of the original quantum-enhanced Monte Carlo  algorithm holds true for the considered problem class;
(ii) the approximate version, using a classical emulator, is able to partially retain this advantage; and
(iii) the problem instance size is sufficiently large such that the runtime prefactor does not overshadow the scaling improvement.
We anticipate that the best emulator for this specific task will be a classical algorithm that optimally balances accuracy and computational cost, even sacrificing the former for the latter. For instance, in Ref.\cite{gangat2025linear} a truncated evaluation of tensor networks using limited-size blocks enabled the approximate simulation of short-range spin glasses up to sizes of $50\times50\times50$.

Finally, we note that the quantum-enhanced Monte Carlo algorithm, due to its peculiar hybrid nature, is particularly robust to various forms of noise, from hardware-related to algorithmic. Noise can only impact its efficiency but will never affect the accuracy of the final results, as it solely affects the proposal step.
Therefore, it stands out as one of the best candidates to be executed either on noisy hardware or through its quantum-inspired classical approximation.\\

\textit{Acknowledgments.} 
We acknowledge useful discussions with Alev Orfi, Dries Sels, and Marina Marinkovic.
G.M. acknowledges financial support from the Swiss National Science Foundation (Grant No. PCEFP2\_203455).

\appendix

\section{Phase transition in quantum spin glass}
\label{s:pt}
\begin{figure}[!b]
    \centering
    \includegraphics{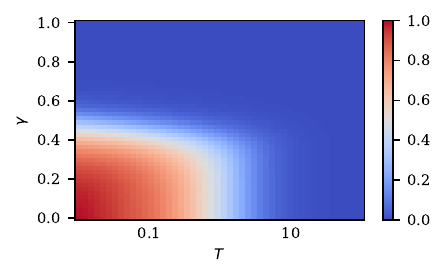}
    \caption{Phase diagram of a quantum spin glass with respect to the field strength $\gamma$ and the temperature $T$ for an $n=9$ system. Each grid point is an average over 100 random instances. The spin glass phase is in red and the paramagnetic phase is in blue.}
    \label{fig:phase_diagram}
\end{figure}

The order parameter used is the Edwards-Anderson parameter~\cite{castellani2005spin}
\begin{equation}
    q = \frac{1}{n}\sum_{i=1}^{n}\overline{\langle\sigma_i\rangle^2},
\end{equation}
where $\langle \cdots \rangle$ denotes the thermal average and the overbar indicates the disorder average, i.e., the average over different spin-glass realizations. Figure~\ref{fig:phase_diagram} presents the phase diagram for $n=9$.

Phase transitions between the spin glass and paramagnetic phases occur along both the $\gamma$ and $T$ dimensions~\cite{de1978stability,young2017stability,yamamoto1987perturbation,mukherjee2015classical}. These transitions are characterized by their dominant factors: quantum fluctuations in the case of $\gamma$ and thermal fluctuations for $T$. Here we focus on the phase transition in $\gamma$, aiming to determine the critical value $\gamma_c$.

\begin{figure}[!b]
    \centering
    \includegraphics{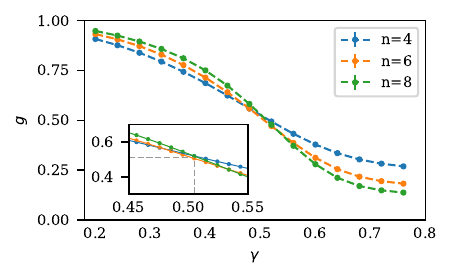}
    \caption{Binder cumulant $g$ as a function of field strength $\gamma$ for different system sizes: $n=4$ (blue), $n=6$ (orange), and $n=8$ (green). The intersection point indicates the critical value $\gamma_c=0.50\pm0.02$, where the phase transition occurs. The uncertainty is from the finite step size.}
    \label{fig:binder_cumulant}
\end{figure}

In finite systems, phase transitions are not sharp, and order parameters exhibit finite-size scaling. We use the Binder cumulant analysis to identify the critical point. The Binder cumulant is defined in terms of the moments of the order parameter~\cite{parisi1983orderparameters}
\begin{equation}
    g = \frac{1}{2} \left[ 3 - \overline{\left( \frac{q^{(4)}}{(q^{(2))^2}}\right)} \right],
\end{equation}
where
\begin{equation}
    q^{(k)} = \frac{1}{n^k} \sum_{i_1 \ldots i_k} \overline{\langle \sigma_{i_1} \cdots \sigma_{i_k} \rangle^2}.
\end{equation}
By plotting $g$ against $\gamma$ for various system sizes, we can estimate $\gamma_c$ from the intersection point of these curves. Figure~\ref{fig:binder_cumulant} illustrates this analysis, from which we determine that the phase transition in $\gamma$ occurs at $\gamma_c = 0.50\pm0.02$ at zero temperature.

\section{Bayesian optimization of time-dependent schedules}
~\label{sec:bayesian_optimization}

In Bayesian optimization, the goal is to maximize a black-box function $f$ defined over a parameter space $D_\theta \subset \mathbb{R}^m$. Bayesian optimization constructs a surrogate model to approximate the objective function, often using Gaussian processes (GPs). The GP model $\tilde{f}(\theta)$ is characterized by a mean function $\mu(\theta)$ and a kernel function $k(\theta, \theta')$, which captures the correlation between points. Here we employ the Matérn 5/2 kernel~\cite{wilson2018maximizing}. By conditioning the GP on observed data, the model updates its predictions, incorporating new information. The updated GP is then used to determine the next point to probe by maximizing the acquisition function $\phi(\theta; \kappa)$. Here we use the upper confidence bound, defined as
\begin{equation}
\phi(\theta; \kappa) = \mu(\theta) + \kappa \cdot \sigma(\theta),
\end{equation}
where $\sigma(\theta)$ represents the predicted uncertainty. The acquisition function balances exploration (sampling regions with high uncertainty) and exploitation (sampling regions with high predicted values). As the optimization progresses, the parameter $\kappa$ is decreased to gradually shift the focus from exploration to exploitation. This iterative process of updating the surrogate model and selecting new points based on the acquisition function continues until convergence. Bayesian optimization has recently been employed to design quantum annealing schedules~\cite{finvzgar2024designing} and to optimize the parameters of the QAOA~\cite{tibaldi2023bayesian}.

In our case, the objective is to optimize the shape of $\gamma(t)$ to produce a proposal distribution with the maximum absolute spectral gap. We fix the total evolution time at $\tau=10$ and define the dimensionless fraction $s(t) = \frac{t}{\tau}$ of the total evolution time. Initially, $\gamma(s(t))$ is parametrized by selecting five equidistant points. Intermediate values are then obtained using cubic interpolation, resulting in a piecewise cubic schedule. Additionally, we enforce $\gamma(0) = \gamma(\tau) = 0$ to ensure that the schedule begins with a classical Hamiltonian, thereby allowing the perturbation to be switched on gradually.

The cost function in our optimization problem takes a five-dimensional vector $\theta$, evaluates the proposal distribution by integrating the time-dependent Schrödinger equation, and calculates the absolute spectral gap of the transition matrix, which serves as our figure of merit.

\begin{figure*}
    \centering
    \includegraphics[width=0.9\textwidth]{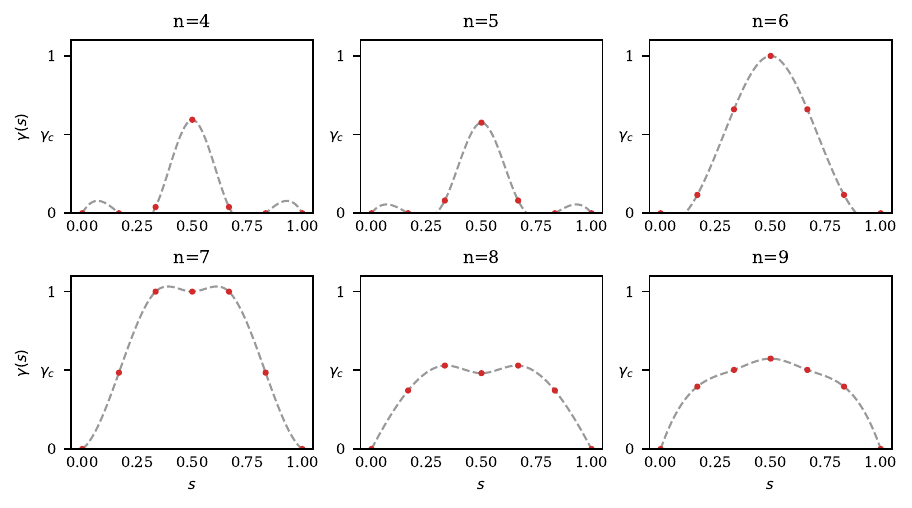}
    \caption{Optimized schedules $\gamma(s)$ for different system sizes obtained using BO: (a) $n=4$, (b) $n=5$, (c) $n=6$, (d) $n=7$, (e) $n=8$, and (f) $n=9$.The x axis represents the dimensionless fraction $s = \frac{t}{\tau}$. The schedules are constrained to be symmetric around the middle point. The critical value $\gamma_c$ is indicated for reference.}
    \label{fig:bo_optimized_schedules}
\end{figure*}

We perform the aforementioned optimization for system sizes ranging from $n=4$ to $9$. The resulting optimized schedules are depicted in Fig.~\ref{fig:bo_optimized_schedules}. Remarkably, we observe that the optimizer converges to similar schedule shapes for system sizes $n \in [4,5]$, $n \in [6,7]$, and $n \in [8,9]$, respectively. For $n \in [6,7]$, the schedule ramps up the transverse field sharply to $\gamma \approx 1$, where the Hamiltonian is dominated by the mixing term. In contrast, for $n \in [8,9]$, the optimized schedule resembles the time-independent case. Here $\gamma(t)$ increases to approximately its critical value, maintains an almost constant field strength, and then symmetrically decreases towards the end.

\section{Violation of Detailed Balance}
\label{app:detailbalance}

The Trotter error does not impact the symmetry condition, because the $U$ and its transpose $U^T$ only differ by a phase. We can verify this by expressing the second-order Trotterized unitary in terms of its transpose and by noting that the $e^{-iH_Z}$ is a diagonal matrix in the computational basis: 

\begin{equation}
    U(t)_{\text{2nd}} = e^{-2iH_Z dt}[U(t)_{\text{2nd}}]^T e^{+2iH_Z dt} = [U(t)_{\text{2nd}}]^T
    \label{eq:first_second_order_trotter_difference}
\end{equation} 

The $U(t)_{\text{2nd}}$ is defined in Eq.~\ref{eq:2nd_Order_Trotter_Def}. In the verification, we used the fact that $H_Z = H_Z^T$ and $H_X = H_X^T$ are symmetric and that for matrices $A,B$ and $C$ the identities $(ABC)^T=C^TB^TA^T$, $(A^n)^T=(A^T)^n$, and $(A+B)^T = A^T+B^T$ hold and therefore $(e^A)^T=e^{(A^T)}$.

\section{Impact of approximation errors on the scaling}

\begin{figure}[!b]
    \centering
    \includegraphics{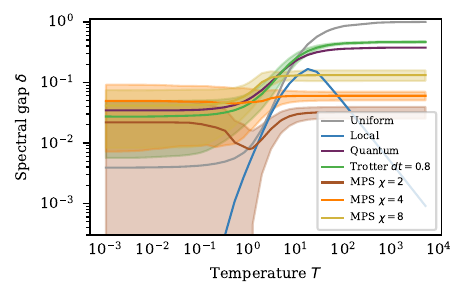}
    \caption{Spectral gap $\delta$ for different temperatures $T$ at fixed system size $n=8$ for different proposal strategies, with $\gamma=0.45$ and $t=12$.  Trotter $dt = 0.8$ refers to the Trotterized quantum proposal. MPS $\chi$ refers to using TEBD with the MPS to perform the time evolution with bond dimension $\chi$. Each data point is averaged over 100 instances. Bond dimension $\chi=16$ would result in the exact representation of the quantum Trotter proposal.}
    \label{fig:MPSgapscalingT}
\end{figure}

Figure ~\ref{fig:MPSgapscaling} in Sec.~\ref{subs:Scaling of the approximate MPS-QEMC} depicts the scaling of the spectral gap in system size for different proposal strategies at fixed temperature $T=1$. Figure ~\ref{fig:MPSgapscalingT} depicts the scaling of the spectral gap in temperature $T$ for different proposal strategies at fixed system size $n=8$. We observe that the performance of the MPS proposals varies greatly with the bond dimension $\chi$.

\section{Details of the Tensor-network calculations}
\label{app:Details of the Tensor-network calculations}

This appendix details the implementation of the quantum-inspired proposal step in Sec.~\ref{ss:mps}. The numerical simulations are implemented in the \textsc{Julia} programming language using the \textsc{itensor.jl} library~\cite{fishman2022itensor}. The proposal requires us to implement the $| \psi(t) \rangle = U | \psi(0)\rangle$. Given the Hamiltonian $H=(1-\gamma)\alpha H_c + \gamma H_{mix} = H_Z + H_X$, we split $U(t) = e^{-iHt}$ into the product 

\begin{equation}
    U(t)_{\text {2nd order }}=e^{+i H_Z (dt/2)}\left[U(t)_{1 \text {st order}}\right] e^{-i H_Z (dt/2)}
    \label{eq:2nd_Order_Trotter_Def}
\end{equation}

to get the second-order Trotterized unitary where 

$$U(t)_{\text{1st order}}=e^{-i H_z^{(1)} d t} e^{-i H_x^{(1)} d t} \cdots e^{-i H_z^{(m)} d t} e^{-i H_X^{(m)} dt},$$

with $t = m dt$. The first and second-order Trotterized unitaries only differ by a phase, which is irrelevant for the associated probability distribution.

We further divide the Trotterized unitaries into one-site and two-site gates. Assuming a complete (or fully connected) graph of $n$ spins, there are $n$ vertices and $n(n-1) / 2$ edges. Therefore, we have to construct $n$ $Z$ gates, $n$ $X$ gates, and $n(n-1) / 2$ $ZZ$ gates. The $X$ and $Z$ gates are single-site gates of the form

$$
G\left(Z_{\sigma_j}\right)=e^{-i d t \gamma Z_{\sigma_j}}, \quad G\left(X_{\sigma_j}\right)=e^{-i d t(1-\gamma) X_{\sigma_j}}
$$

and are represented as rank-2 tensors (two indices) where each index is of dimension $d=2$ for the spin-$1/2$ particles. The $ZZ$ gates are of the form

$$
G\left(Z Z_{\sigma_j, \sigma_k}\right)=e^{-i dt \gamma Z_{\sigma_j} Z_{\sigma_k}}
$$

and are represented as rank-4 tensors where each index is of dimension $d=2$ as well. The one-site and two-site gates are the building blocks required for the TEBD of the MPSs. The procedure can be separated into the following steps:

\begin{enumerate}
    \item[\textit{(i)}]  \textit{System setup}. The underlying (complete) graph of spins and their tensor indices is created. The graph is mapped to an MPS. The MPS is initialized. 
    \item[\textit{(ii)}] \textit{Gate construction}. The unitary evolution is Trotterized and split into one-site gates (related to the $X$ or $Z$ terms in the Hamiltonian) and two-site gates (related to the $ZZ$ terms in the Hamiltonian). The gates represent an evolution step and are computed as matrix exponentials of the Hamiltonian. 
    \item[\textit{(iii)}]  \textit{Gate application}. The time evolution of the state is done by applying the gates to the MPS. 
    \item[\textit{(iv)}] \textit{Normalization}. After each time step the MPS is normalized. 
    \item[\textit{(v)}]  \textit{Sampling}. The state is sampled after concluding the time evolution, which results in the new proposal step. 
\end{enumerate}

\section{Memory and Computational Cost Estimates of TEBD for MPS}
\label{app:Memory and Computational Cost Estimates of TEBD for MPS}

This appendix details the assumptions and derivations in the memory and computational cost estimates for the local, uniform, and MPS proposal strategies. The estimates serve as a basis for estimates of the time needed to perform an individual Markov chain step [$t_c(n)$ for local and uniform, and $t_{qi}$ for the QIMC] in Sec.~\ref{s:Threshold for practical quantum inspired advantage}.

The local proposal strategy consists of randomly drawing one of $n$ spins and flipping it. This corresponds to storing $n$ complex coefficients and a computational complexity of $O(1)$. The uniform proposal also has to store the $n$ coefficients but draws $n$ spins randomly, resulting in a computational complexity of $O(n)$. Compared to the local and uniform proposal strategies, the MPS proposal strategy requires vastly more time and memory resources. In the following, we use $d$ for the dimension of the local spins, assume a fully connected spin graph with $n$ vertices and $n(n-1)/2$ edges, and look at the TEBD of the MPS as described in Appendix~\ref{app:Details of the Tensor-network calculations}. 

The dominating memory costs are the storage of the matrix product state with $O(n d \chi^2)$ and the storage of the application gates. We need to store $2n$ single-site gates (for the $X$ and $Z$ terms in the Hamiltonian) and $n(n-1)/2 \propto n^2$ two-site gates (for the $ZZ$ terms in the Hamiltonian). The two-site gates are the dominating term and are stored as rank-4 tensors where each index is of dimension $d$ resulting in an $O(n^2 d^4)$ contribution. Therefore, the memory cost can be approximated by $O\left(n d \chi^2+n^2 d^4\right)$. For instance, using 16 GB of memory, we can store an MPS and the associated gates for $n=16$ with the exact bond dimension of $\chi=256$ or $n=100$ with $\chi=95$. In both cases, we use complex coefficients and the single-precision floating-point format (float32). 

In terms of computational time, we can neglect the system setup, the gate construction, the one-site gate applications, the normalization, and the sampling, because the dominating factor is the two-site gate application.

The gate application represents the time evolution of the state. For each Trotter step, we have to apply all one- and two-site gates once. In total, there are $m$ Trotter steps resulting in $t=m d t$. The application of a one-site gate can be done without incurring an error and without increasing the memory required. The number of contractions needed to apply the one-site gate to the MPS is $O\left(d^2 \chi^2\right)$.

The application of a two-site gate is more complicated. We have to distinguish between sites that are adjacent in the MPS and sites that are non adjacent. For adjacent sites, the gate application consists of contracting all indices in an efficient order, reshaping the indices, applying a singular value decomposition (SVD), and performing a truncation to the desired bond dimension or accuracy. The contraction of all indices requires $O\left(d^4 \chi^2+\chi^3 d^3\right)=O\left(\chi^3 d^3\right)$ operations. The SVD is performed in $O\left(\chi^3 d^3\right)$ operations. If we do not truncate the bond dimension between the two sites, the gate application grows the bond dimension from $\chi$ to $\chi d$.

If the sites are not adjacent, there are different ways of applying the two-site gate. In the following, we assume that we have two sites $i$ and $j$ that are separated by $L=j-i+1$. The brute-force approach would require us to contract all intermediate bond indices between sites $i$ and $j$, reshuffle the indices, apply the two-site gate and decompose the resulting tensor back into MPS form. This is really inefficient.

To get a better estimate for the number of contractions necessary to perform a non adjacent two-site gate we follow the arguments provided in~\cite{shi2006classical, paeckel2019time} for long-range interactions in one-dimensional systems. The underlying idea is to use \textsc{swap} gates that shuffle the indices in the MPS. This encompasses swapping indices to create adjacent sites, applying the gate, and then swapping back the indices to their original positions. If the bond dimensions are not truncated, the bond dimensions between the two sites increase by $d$ or $d^2$.

Reshuffling the indices of a rank-$a$ tensor, where each index is of dimension $b$, requires $O(b^a)$ operations. Reshuffling the physical indices of two adjacent MPS sites requires us to first contract the two adjacent indices to create a combined tensor $O(d^2 \chi^3)$, then reshuffle the indices of the new tensor $O(d^2\chi^2)$, and finally apply an SVD to get back to the MPS form $O(d^3 \chi^2)$. Overall, one \textsc{swap} gate requires $O(d^2 \chi^3 + d^2\chi^2 + d^3\chi^3)= O(d^3\chi^3)$ operations.

Different two-site gates require different numbers of \textsc{swap} gates depending on their adjacency. Out of the $n(n-1)/2$ two-site gates, $n-1$ require zero \textsc{swap} gates, $n-2$ require two \textsc{swap} gates, $n-3$ require four \textsc{swap} gates ending with one gate requiring $n-2$ \textsc{swap} gates. Hence, we have to perform a total of

\begin{equation}
    2 \sum_{a=1}^{n-1} (n-a)(a-1) = \frac{n^3}{3} +  n^2 + \frac{2}{3} n 
\end{equation}

\textsc{swap} operations for one Trotter step. Each \textsc{swap} gate is $O(d^3 \chi^3)$. Therefore, applying all two-site gates for one Trotter step requires operations of 

\begin{equation}
    O \left( \left(\frac{n^3}{3} + n^2 + \frac{2}{3}n \right) d^3 \chi^3 \right) =  O \left(n^3 d^3 \chi^3 \right), 
\end{equation} 

which is the largest contributing factor in the application of the two-site gate application. In addition, if we do not truncate the bond dimension after each gate application, the maximum bond dimension $\chi=2^{n/2}$ is reached after applying the first $n$ two-site gates, i.e., the two-site gates connecting the leftmost index with all the other indices. Accounting for the $m$ Trotter steps, this results in contractions of $O(m n^3 d^3 \chi^3)$, as summarized in Table \ref{tab:computational_cost}.

\bibliography{biblio.bib}

\end{document}